\documentclass{article}
\usepackage[margin=2cm]{geometry}

\usepackage{float}

\usepackage{changepage}

\usepackage[utf8]{inputenc}

\usepackage{textcomp,marvosym}

\usepackage{fixltx2e}

\usepackage{amsmath,amssymb}

\usepackage{cite}

\usepackage{nameref,hyperref}

\usepackage{microtype}
\DisableLigatures[f]{encoding = *, family = * }

\usepackage{rotating}

\usepackage[aboveskip=1pt,labelfont=bf,labelsep=period,justification=raggedright,singlelinecheck=off]{caption}


\usepackage{natbib}

\DeclareMathOperator{\Cov}{Cov}

\DeclareMathOperator{\E}{\mathbb{E}}

\newcommand{\Covb}[2]{\ensuremath{\Cov\!\left[#1,#2\right]}}
\newcommand{\Eb}[1]{\ensuremath{\E\!\left[#1\right]}}

\newcommand{\norm}[1]{\| #1 \|}

\usepackage{ragged2e}

\title{Indirect Evidence of Network Effects in a System of Cities}
\date{}

\author{Raimbault Juste\textsuperscript{1,2,*}\\
\textsuperscript{1}UMR CNRS 8504 G{\'e}ographie-cit{\'e}s, Paris, France\\
\textsuperscript{2}UMR-T 9403 IFSTTAR LVMT, Champs-sur-Marne, France\\
Email : juste.raimbault@polytechnique.edu}

\begin{document}

\maketitle

\begin{abstract}
We describe a simple spatial model of urban growth for systems of cities at the macroscopic scale, which combines direct interaction between cities and an indirect effect of physical network flows as population growth drivers. The model is parametrized on population data for the French system of cities between 1831 and 1999, which strong non-stationarity in correlation patterns suggest to apply the model on local time windows. The corresponding calibration of the model using genetic algorithms provide the evolution of interaction processes and network effects in time. Furthermore, the fit improvement when adding network module appears effective when controlling for additional parameters, what confirms the ability of the model to unveil network effects in the system of cities.
\end{abstract}


\maketitle


\section*{Introduction}

Cities are paradoxically unsustainable and source of negative externalities, but also the best chance to reach sustainability and resilience to climate change~\citep{glaeser2011triumph}. The dynamics of urban systems at a macroscopic scale, and more precisely drivers of urban growth, are crucial processes that need to be understood to meet these potentialities. A better knowledge of how cities differentiate, interact and grow is thus a relevant topic both from a theoretical perspective and for policy applications. Indeed, \cite{pumain2009innovation} suggest that cities are incubators of social change, their fate being closely linked to the one of societies. 

Various disciplines have studied models of urban growth with different objectives and taking diverse aspects into account. For example, economics are still reluctant to include spatial interactions in models~\citep{krugman1998space} but are extremely detailed on market processes, even for models in economic geography. On the other hand, geography focuses more on territorial specificities and interactions in space but will have more difficulties to produce general conclusions. The example of this two disciplines and their misunderstandings \citep{marchionni2004geographical} shows how it is difficult to make bridges, how it needs exceptional efforts to translate from one to the other (as P. Hall did for Von Thunen work~\citep{taylor2016polymath}), and therefore how it is far from evident to grasp the complexity of urban systems in an integrated way.

The simplest model to explain urban growth, the Gibrat model, assumes random growth rates for cities. It has been shown by~\cite{gabaix1999zipf} that this model asymptotically produce the expected rank-size law (Zipf's law) for system of cities, which is considered as one of the most regular stylized facts, at least in its generalized scaling law formulation~\citep{nitsch2005zipf}. Explaining urban scaling laws is closely related to the understanding of urban growth, as \cite{bettencourt2008large} suggest that these reflect underlying universal processes and that all cities are scaled version of each other. This approach however does not reflect the complex relation between economic agents for which~\cite{storper2009rethinking} advocate. Using a bottom-up reconstruction of urban areas with dynamical microscopic population data, \cite{rozenfeld2008laws} show indeed that positive deviations to the rank-size law systematically exist, and that these must be an effect of spatial interactions between urban areas. 

Complexity approaches are good candidates to integrate these interactions into models. \cite{andersson2006complex} introduce for example a model of urban economy as a growing complex network of relations. The Evolutive Urban Theory, introduced by~\cite{pumain1997pour}, focuses on cities as co-evolving entities and produces explanations for growth at the scale of the system of cities. \cite{pumain2006evolutionary} show that scaling laws could be a consequence of functional differentiations and of the diffusion of innovation between cities. The positioning regarding universality of laws is more moderate than theories of scaling \citep{west2017scale}, as \cite{pumain2012urban} highlights that ergodicity can difficultly be assumed in the frame of complex territorial systems. One crucial feature of this paradigm is the importance of interactions between agents, generally the cities, to produce the emergent patterns at the scale of the system. \cite{pumain2013theoretical} investigate the advantages of agent-based models in comparison to more classical equation systems, and this methodological aspect is in accordance with the theoretical positioning, as it allows to take into account the heterogeneity of possible interactions, the geographical particularities, and to naturally include emergence between levels and render multi-scale patterns.

In this paper we aim at exploring further the assumption, central to Pumain's Evolutive Urban Theory, that spatial interactions between cities are significant drivers of their growth. More precisely, we consider both abstract interactions and flow interactions mediated through the physical networks, mainly the transportation networks. We extend existing models accordingly. Our contribution is twofold: (i) we show that very basic interaction models based on population only can be fitted to empirical data and that adjusted parameter values can be directly interpreted; and (ii) we introduce a novel methodology to quantify overfitting in models of simulation, as an extension of Information Criteria for statistical models, which application to our calibrated models confirms that fit improvement is not only due to additional parameters, but that the extended model effectively capture more information on system processes. This unveils network effects in an indirect way. We first review modeling approaches to urban growth based on spatial interactions.

\subsection*{Urban growth and spatial interactions}

First of all, we must precise that we consider only models at the macroscopic scale, ruling out the numerous and rich approaches at the mesoscopic scale, that include for exemple cellular automata models, models of urban morphogenesis or land-use change models. Some include explicitly spatial interactions: \cite{rybski2013distance} show that a gravity-based aggregation of population in a lattice produces realistic metropolitan patterns. Such kind of models operate however at an other level of detail than the one we consider. We also naturally rule out economic models that do not explicitly include spatial interactions.

Several models of urban growth at the macro scale have insisted on the role of space and spatial interactions. \cite{frasco2014spatially} introduce a spatialized model of social network growth that provides a good agreement with observed data both for city size hierarchy and the relation between population and density. \cite{bretagnolle2000long} propose a spatial extension of the Gibrat model. The gravity-based interaction model that~\cite{sanders1992systeme} uses to apply concepts of synergetics to cities is also close to this idea of interdependent urban growth, contained physically in the phenomenon of migration between cities. 

A more refined extension with economic cycles and innovation waves is developed by~\cite{favaro2011gibrat}, yielding a version of the core of Simpop models~\citep{pumain2012multi} in terms of dynamical systems. This family of models have started with a toy-model based on economic interactions between cities as agents, that yield hierarchical patterns at the scale of the system~\citep{sanders1997simpop}. The Simpop2 model, still based on spatial interaction for commercial exchanges, including successive innovation waves, unveils structural differences between the European and the US urban systems~\citep{bretagnolle2010comparer}. The SimpopLocal model~\citep{pumain2017simpoplocal} simulates the emergence of initial settlement patterns. The Marius model~\citep{cottineau2014evolution} couples population and economic growth with interactions between cities, allowing to accurately reproduce real trajectories on the former Soviet Union after calibration with processes multi-modeling.

\subsection*{Urban Growth and Transportation Networks}

Under similar assumptions of previously reviewed models, the inclusion of transportation networks has been rarely pursued, contrary to the mesoscopic scale at which relations between networks and territories have been widely studied by LUTI models for example~\citep{chang2006models}. Transportation network growth models~\citep{xie2009modeling}, prolific in economics and physics, are designed to reproduce network topology and do not focus mainly on the influence of networks on cities. \cite{bigotte2010integrated} study an optimization model for network design combining the effects of urban hierarchy and of transportation network hierarchy. \cite{baptiste1999interactions} models the dynamical interplay between network links capacity and city growth on a subset of the French city system. The SimpopNet model~\citep{schmitt2014modelisation} goes a step further in modeling the co-evolution between cities and transportation networks, as it allows new network links to be created in time. These examples shows the difficulty of coupling these two aspects of urban systems in models of growth, and we will for this reason take into account network effects in a simplified way as detailed further.

The rest of this paper is organized as follows: our model is introduced and formally described in the next section; we then describe results obtained through exploration and calibration of the model on data for French cities, in particular the unveiling of network effects significantly influencing growth processes thanks to a novel methodology introduced. We finally discuss implications of these results.

\section*{Model Description}

\subsection*{Rationale}

Some confusion may arise when surveying stochastic and deterministic models of urban growth. To what extent is a proposed model ``complex'' and is the simulation of stochasticity necessary ? Concerning the Gibrat model and most of its extensions, independence assumptions and linearity produce a totally predictable behavior, which is thus not complex in the sense of exhibiting emergence, in the sense of weak emergence~\citep{bedau2002downward}. In particular, the full distribution of random growth models can be analytically determined at any time~\citep{gabaix1999zipf}, and in the case of studying only the first moment, a simple recurrence relation avoids to proceed to any Monte-Carlo simulation. Under these assumptions, it is natural to work with a deterministic model, as it is done for example for the Marius model~\citep{cottineau2014evolution}. We will work with that hypothesis, capturing complexity through non-linearity.

We work on simple territorial systems assumed as regional city systems, in which cities are basic entities. The time scale corresponds to the characteristic scale associated to this spatial scale, i.e. around one or two centuries. Spatial interactions are captured through gravity-type interactions, this simple formulation having the advantage of being simple and of capturing the first law of Tobler, namely that interaction strength fades with distance. Other approaches introduced recently perform similarly at this scale~\citep{masucci2013gravity}.

\subsection*{Model description}

We consider a deterministic extension of the Gibrat model, what is equivalent to consider only expectancies in time. Let $\vec{P}(t)=(P_i(t))_{1\leq i\leq n}$ be the population of cities in time. Under Gibrat independence assumptions, we have $\Covb{P_i(t)}{P_j(t)}=0$. A linear extended version would write $\vec{P}(t+1)=\mathbf{R}\cdot \vec{P}(t)$ where $\mathbf{R}$ is an independent random matrix of growth rates (identity in the original case). This yields directly thanks to the independence assumption that $\Eb{\vec{P}(t+1)}=\Eb{\mathbf{R}}\cdot\Eb{\vec{P}}(t)$. We generalize this linear relation to a non-linear relation that allows to be more consistent with model interpretation and more flexible. Denoting $\vec{\mu}(t)=\Eb{\vec{P}(t)}$, we take

\begin{equation}
	\vec{\mu}(t+1)=\Delta t\cdot f(\vec{\mu}(t))
\end{equation}

Note that in that case, stochastic and deterministic versions are not equivalent anymore, precisely because of the non-linearity, but we stick to a simple deterministic version for the sake of simplicity. The specification of the interdependent growth rate is thus given by

\begin{equation}
f(\vec{\mu}) = r_0\cdot \mathbf{Id}\cdot \vec{\mu} + \mathbf{G}\left(\vec{\mu}\right)\cdot \vec{1} + \vec{N}\left(\vec{\mu}\right)
\end{equation}

where $\vec{1}$ is the column vector full of ones, and $\mathbf{G} = G_{ij} = w_G\cdot \frac{V_{ij}}{<V_{ij}>}$ such that the interaction potential $V_{ij}$ follows a gravity-type expression given by, with $d_{ij}$ distance between $i$ and $j$ (euclidian or network distance),

\begin{equation}
V_{ij} = \left(\frac{\mu_i\mu_j}{\left(\sum_k{\mu_k}\right)^2}\right)^{\gamma_G}\cdot \exp{\left(-d_{ij}/d_G\right)}
\end{equation}

The network effect term $\vec{N}$ is given by $N_{i} = w_N \cdot \frac{W_i}{<W_i>}$ where the network flow potential $W_i$ reads

\begin{equation}
W_{i} = \sum_{k < l} \left(\frac{\mu_k\mu_l}{\left(\sum_j\mu_j\right)^2}\right)^{\gamma_N} \cdot \exp{\left(-d_{kl,i}/d_N\right)}
\end{equation}

where $d_{kl,i}$ is the distance of city $i$ to the shortest path between $k,l$ computed in the geographical space, which can be through a transportation network or in an impedance field of the euclidian network. All seven model parameters are detailed below.

The first component corresponds to the pure Gibrat model, that we obtain by setting the weights $w_G = w_N = 0$. The second component captures direct interdependencies between cities, under the form of a separable gravity potential such as the one used in~\cite{sanders1992systeme}. The third term is aimed at capturing network effects by expressing a feedback of network flow between cities $k,l$ on the city $i$. Intuitively, a demographic and economic flow physically transiting through a city or in its surroundings is expected to influence its development (through intermediate stops e.g.), this effect being of course dependent on the transportation mode since a high speed line with few stops will skip most of the traversed territories. Note that we don't use exactly gravity flows in the network term, since there is no decay of interactions generating flows with distance, but a decay of the effect of the flow as a distance to the network: it is equivalent to assuming a long-range use of the network on average in time, and is this way complementary to the first gravity term.

\subsection*{Model Parameter Space}

We give in Table~\ref{tab:parameters} the description of model parameters, detailing the associated processes and parameter ranges. Both direct interaction and second order network flows effect have the same structure, namely separability between the effect of distance and the influence of population, an exponential decay parameter and a hierarchy parameter expressing the inequality of contribution depending on relative sizes of cities: the highest the exponent, the more contribution of smaller cities will be negligible regarding larger cities.

We propose to interpret the distance decay parameter the following way. Let fix an arbitrary fraction $\alpha$ and typical spatial ranges for a local urban system $d_L$ and for a long range urban system $d_R$, consider a city $i$ and two neighbors $j,j'$ with same population $\mu_j=\mu_j'$, at distances $d_L$ and $d_R$ of $i$ respectively. If we want to answer the question to what distance difference is equivalent an attenuation of $\alpha$ of the interaction potential with $i$, we obtain $d_L - d_R = -d_G\cdot \ln \alpha$. Therefore, $d_G$ is exactly the proportionality coefficient answering this intuitive request.

Finally, we will consider only positive weights $w_G$ and $w_N$, to follow empirical observations as detailed below. Numerical values for the weights will be given normalized by number of cities implied in the process, i.e. ${w'}_G = w_G / n$ and ${w'}_N = w_N / (n (n-1) / 2)$.

\begin{table}[ht]
\small\sf\centering
\caption{\textbf{Model parameters summary.} We give the parameters names, notations, associated processes, possible interpretations, typical variation ranges and units.}\label{tab:parameters}
\hspace{-0.5cm}
\begin{tabular}{|l|l|l|l|l|l|}
\hline
Parameter & Notation & Process & Interpretation & Range & Unit\\
\hline
Growth Rate & $r_0$ & Endogenous growth & Growth rate & $\left[ 0,1\right]$ & 1 \\
Gravity weight & $w_G$ & Direct interaction & Max average rate & $\left[ 0,1\right]$ & 1 \\
Gravity gamma & $\gamma_G$ & Direct interaction & Level of hierarchy & $\left[ 0,+\infty\right]$ & 1 \\
Gravity decay & $d_G$ & Direct interaction & Interaction range & $\left[ 0,+\infty\right]$ & km \\
Feedback weight & $w_N$ & Flows effect & Max average rate & $\left[ 0,1\right]$  & 1\\
Feedback gamma & $\gamma_N$ & Flows effect & Level of hierarchy & $\left[ 0,+\infty\right]$ & 1\\
Feedback Decay & $d_N$ & Flows effect & Network effect range & $\left[ 0,+\infty\right]$ & km \\
\hline
\end{tabular}
\end{table}

\subsection*{Data}

Our model is assumed as hybrid as it relies on semi-parametrization on real data. It could be possible to study it as a full toy-model, initial configuration and physical environment being constructed as synthetic data. We however aim at unveiling stylized facts on real data rather than on model behavior in itself, and setup therefore the model from the data we now describe.

\subsubsection*{Population data}

We work with the Pumain-INED historical database for French cities~\citep{pumain1986fichier}, which give populations of \emph{Aires Urbaines} (INSEE definition) at time intervals of 5 years, from 1831 to 1999 (31 observations in time). The latest version of the database integrates Urban Areas, allowing to follow them on long time-period, according to the long time ontology for cities given by \cite{bretagnolle:tel-00459720}, that constructs a functional definition of cities as entities with boundaries evolving in time. We work on the 50 bigger cities in 1999. We furthermore isolate periods of similar length excluding wars, obtaining 9 periods of 20 years\footnote{The time periods are more precisely: 1831-1851, 1841-1861, 1851-1872, 1881-1901, 1891-1911, 1921-1936, 1946-1968, 1962-1982, 1975-1999.} on which adjustment of the model will be done, taking into account non-stationarity in time.

\subsubsection*{Physical flows}

The model takes into account stylized physical flows, in order to be rather independent of network effective shape. These are therefore assumed to follow the geographical shortest path according to terrain slope. This assumption avoids to obtain geographical absurdities such as cities with a difficult access having an overestimated growth rate. Using a 1km resolution Digital Elevation Model, we construct an impedance field, following \cite{collischonn2000direction}, which is given by

\[
Z = \left(1 + \frac{\alpha}{\alpha_0}\right)^{n_0}
\]

where $Z$ is the impedance of links the 1km grid network in which each cell is connected to its eight neighbors. $\alpha$ is the terrain slope computed with elevation difference between the two cells. We take fixed parameter values $\alpha_0 = 3$ (corresponding to approximatively the real world value of a 5\% slope) and $n_0 = 3$ which yield more realistic paths than smaller values\footnote{More precisely, paths between main cities such as Paris-Lyon, Lyon-Marseille, Lyon-Bordeaux, were visually inspected for $\alpha_0 \in \{2,3,4\}$ and $n_0 \in \{2,3\}$, and the parameters giving the most realistic paths were chosen.}.

\subsection*{Indicators of model performance}

We work on an explanatory rather than an exploratory model. Therefore, indicators to evaluate model outputs are not directly linked to intrinsic properties of trajectories or obtained final states, but rather to a distance to the phenomenon we want to explain, i.e. the data. Given real population $p_i(t)$ (historical realizations of $P_i(t)$) and simulated expected populations $\mu_i(t)$ obtained with $\vec{\mu}(t_0) = \vec{p}(t_0)$ on a period of length $T$, we can evaluate two complementary aspects of model performance:

\begin{itemize}
\item Overall model performance, given by logarithm of the mean-square error in space and time
\[
\varepsilon_G = \ln{\left(\frac{1}{T}\sum_t \frac{1}{n} \sum_i \left(p_i (t) - \mu_i (t) \right)^2\right)}
\]
\item Average local model performance, given by the mean-square error on logarithms
\[
\varepsilon_L = \frac{1}{T}\sum_t \frac{1}{n} \sum_i \left(\ln p_i(t) - \ln \mu_i (t)\right)^2
\]
\end{itemize}

Both are actually complementary, as using only $\varepsilon_G$ as it is generally done will focus only on larger cities and give poor results on medium-sized and small cities (for France only Paris will have reasonable fit as it strongly dominates other urban areas and cities). $\varepsilon_L$ allows therefore to take into account model performance in all cities simulated by the model.

\section*{Results}

\subsection*{Stylized facts}

Basic stylized facts can be extracted from such a database, as it has already been widely explored in the literature~\citep{guerin1990150}. We retrieve better fits of log-normal distributions of growth rates at all dates compared to normal fits, and also the fact that growth rates are mainly positive, on the cities we consider and when removing wars.

An interesting feature to look at in relation with our considerations on spatial interactions are correlations between time-series, and more particularly their variation as a function of distance. We consider 50 years overlapping time-windows to have enough temporal observations, finishing respectively in (1881,1906,1931,1962,1999), and estimate on each, for each couple of cities $(i,j)$, the correlation between log-returns $\hat{\rho}_{ij}=\rho\left[\Delta X_i, \Delta X_j\right]$ with a classical Pearson estimator, where
\[\Delta X_i = X_i(t) - X_i(t-1)\]
and
\[X_i(t) = \ln\left(\frac{P_i(t)}{P_i(t_0)}\right)\]
This method, used mainly in econophysics~\citep{mantegna1999introduction}, reveals dynamical interactions without being biased by sizes.

\begin{figure}
\centering
\includegraphics[width=\textwidth]{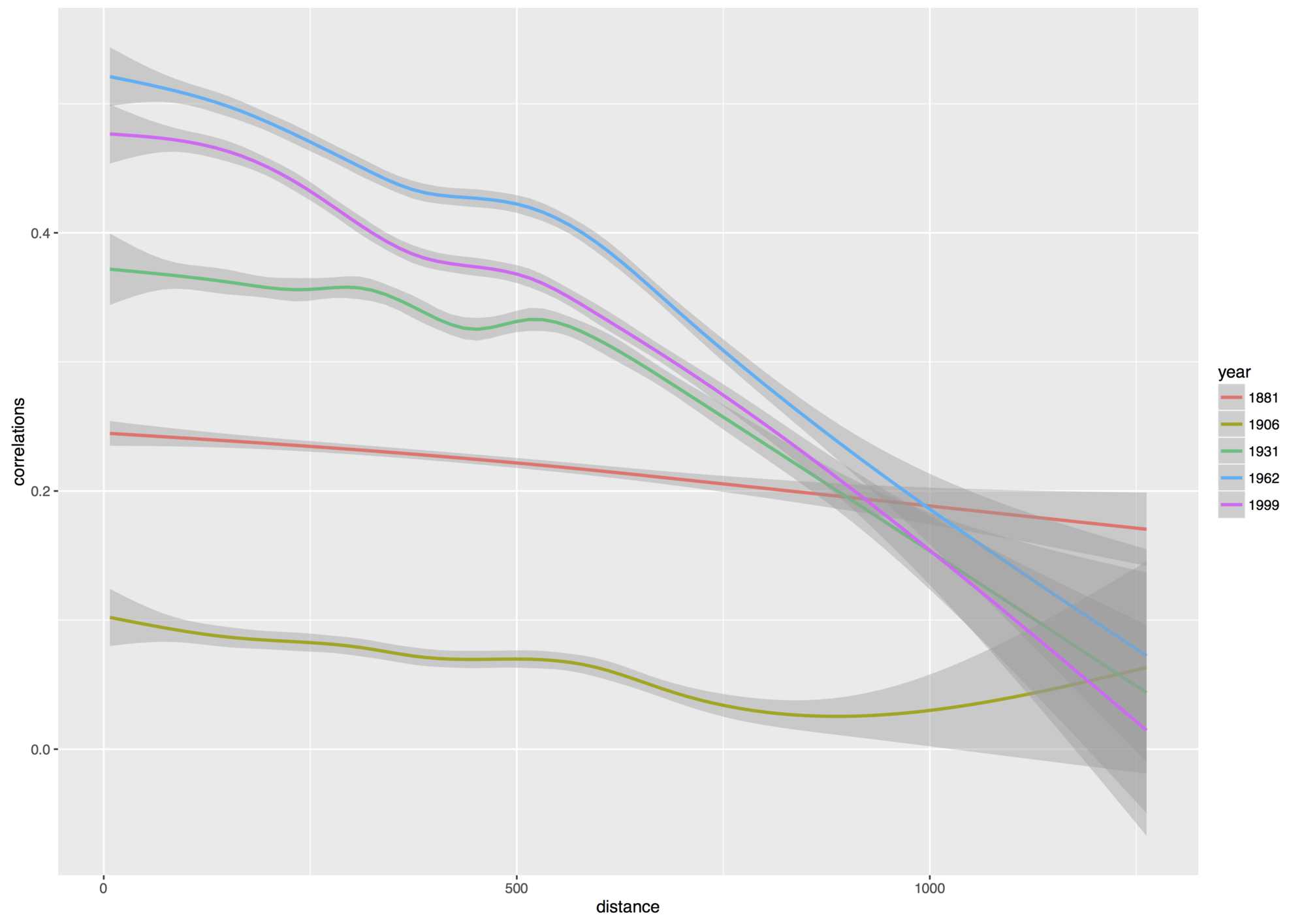}
\caption{\textbf{Time-series correlations as a function of distance.} Solid line correspond to smoothed correlations, computed between each pairs of normalized log-returns of population time-series, on successive periods given by curve color.}
\label{fig:ts-correlations}
\end{figure}

We show in Figure~\ref{fig:ts-correlations} the smoothed correlations curves as a function of distance, for each time period. First of all, the strong differences between each time period confirms the non-stationarity of growth rates over the whole time period, and justifies the use of local fit in time for the model. We can also interpret these patterns in terms of historical events for the system of city and the transportation network. The dynamic of the system begins with a flat correlation in 1881, around 0.2, that could be spurious due to simultaneous similar growth for all cities. It then stays flat but goes to zero, witnessing strong differentiations in growth patterns between 1881 and 1906. After 1931, the effect of the distance is clear with decreasing curves, starting between 0.4 and 0.5. We postulate that this evolution must be partly linked to the evolution of the transportation network: considering railway network for example~\citep{thevenin2013mapping}, the initial overall development may have fostered long range interactions flattening thus the correlation curves, whereas its maturation over time has conducted to the return of more classical interactions decreasing quickly with distance.

\subsection*{Model Exploration}

\paragraph{Implementation}

Data preprocessing, result processing and models profiling are implemented in \texttt{R}. For performances reasons and an easier integration into the OpenMole software for model exploration~\citep{reuillon2013openmole}, a \texttt{scala} version was also developed. The question of a trade-off between implementation performance and interoperability is a typical issue in this kind of model, as a fully blind exploration and calibration can be misleading for further research directions or thematic interpretations. A NetLogo implementation, allowing interactive exploration and dynamical visualization, was also developed for this reason. Source code for models, cleaned raw data, simulation data, and results used in this paper are available on the open repository of the project at \url{https://github.com/JusteRaimbault/InteractionGibrat}.

\begin{figure}
\centering
\includegraphics[width=\textwidth]{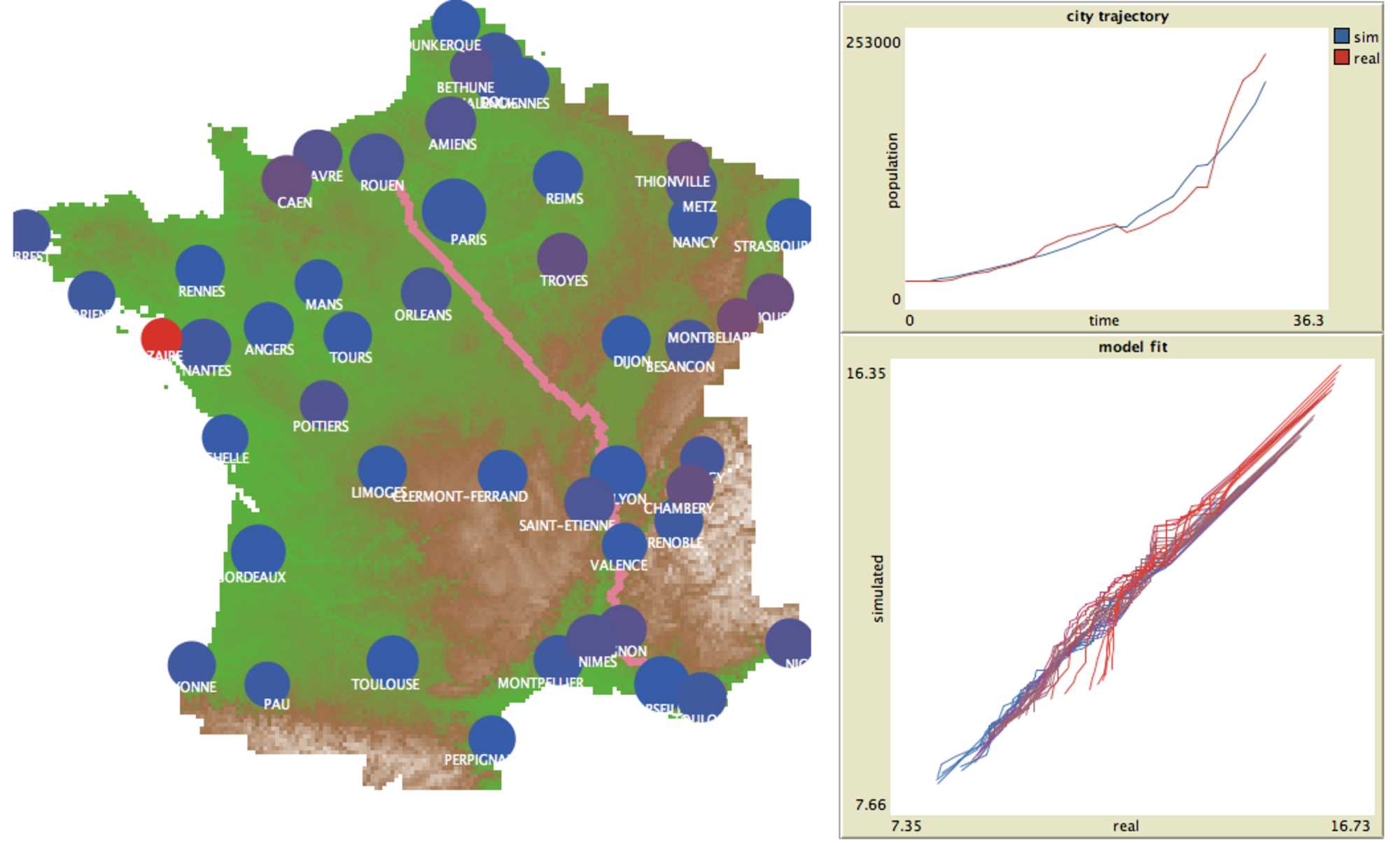}
\caption{\textbf{Example of output of the model.} The graphical interface allows to explore interactively on which cities changes operate after a parameter change, what is necessary to interpret raw calibration results. The map gives adjustment errors by city (color) and their population (size). We illustrate in pink the geographical shortest path between Rouen and Marseille. The plot in top right panel gives in time the trajectory of a selected city, comparing simulated population with real population. The bottom right plot gives for each date the simulated data against real data: the closest the curve is to the diagonal, the better the fit.}
\label{fig:interface}
\end{figure}

We show in Figure~\ref{fig:interface} an example of model output. Cities color give city-level fit error and their size the population. Outliers can therefore easily be spotted (as Saint-Nazaire having the worst fit in the example shown) and possible regional effects identified. We illustrate in pink an example of geographical shortest path, from Rouen to Marseille, which reasonably corresponds to the actual current shortest path by highway. Top right plot shows trajectory in time for a given city, whereas the bottom right plot gives overall fit quality in time, by plotting simulated data against real data. The closest the curve is to the diagonal, the better the fit.

\paragraph{Behavior Patterns}

\begin{figure}
\centering
\includegraphics[width=\textwidth]{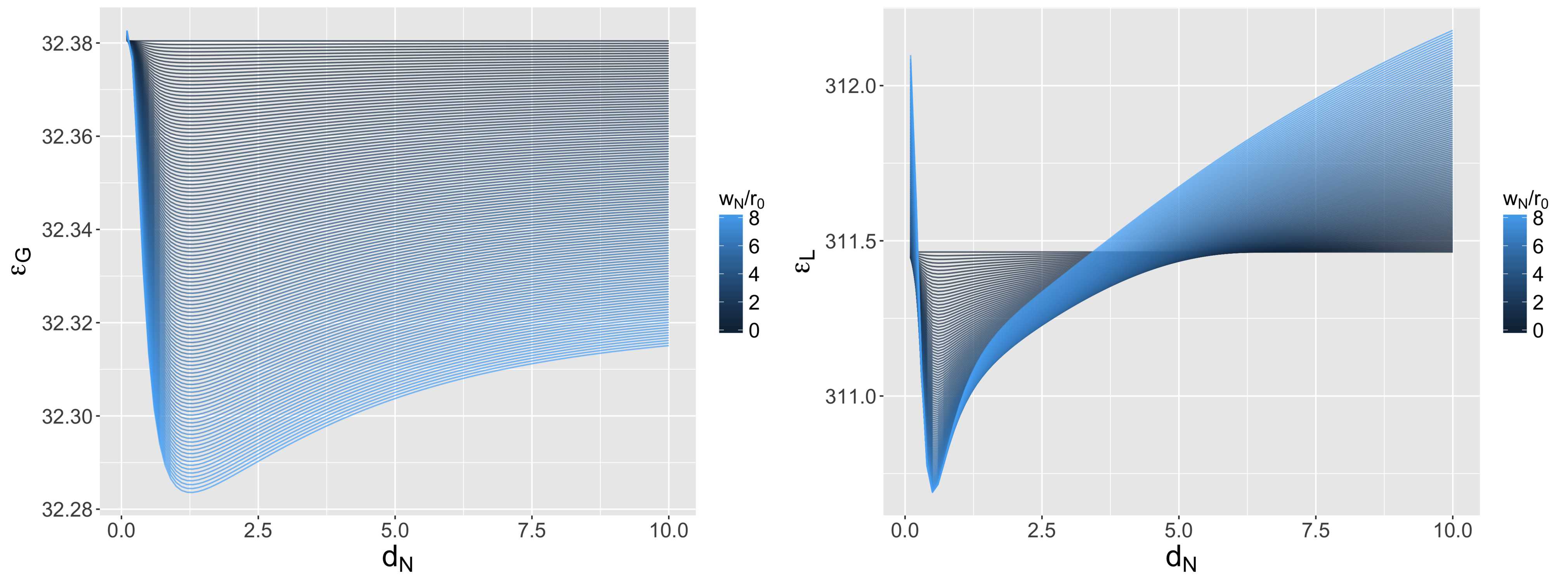}
\caption{\textbf{Evidence of network effects revealed by model exploration.} Left plot gives $\varepsilon_G$ as a function of $d_N$ for varying $r_0/w_N$, at fixed gravity effect and $\gamma_N=3$. Right plot is similar for $\varepsilon_L$}
\label{fig:networkeffects}
\end{figure}

First model explorations, by simply sweeping fixed grids of the parameter space, already suggest the presence of network effects, in the sense that physical flow effectively have an influence on growth rates. We show in Figure~\ref{fig:networkeffects} a configuration of such a case. At fixed gravity parameters and growth rate, we study variations of the parameters $w_N, d_N$ and $\gamma_N$ and the corresponding response of $\varepsilon_G$ and $\varepsilon_L$. At fixed values of $\gamma_N$, we observe similar behaviors of the indicators when $w_N$ and $d_N$ change. The existence of a minimum for both as a function of $d_N$, that becomes stronger when $w_N$ increases, shows that introducing the network feedback terms improves local and global fits compared to the gravity model alone, i.e. that the associated process have potential explanatory power for growth patterns.

\subsection*{Calibrating the Gravity Model}

\begin{figure}
\centering
\includegraphics[width=\textwidth]{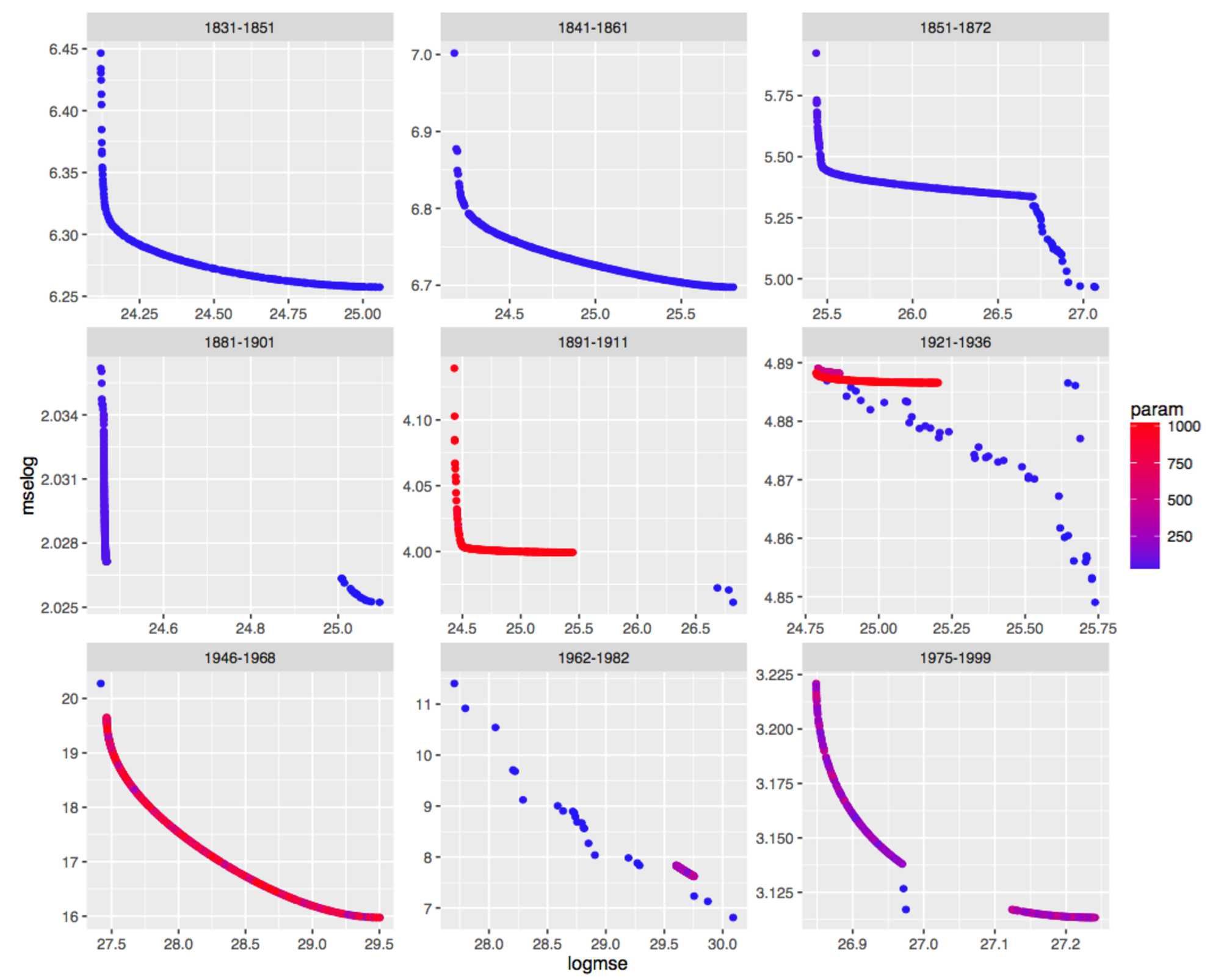}
\caption{\textbf{Calibrating the Gravity Model.} Pareto-front on successive periods. Each panel gives the Pareto front between the two objectives $(\varepsilon_G,\varepsilon_L)$, the dates for the period being given at the top of each. Color level gives the value of distance decay parameter $d_G$.}
\label{fig:gravity-pareto}
\end{figure}

\begin{figure}
\centering
\includegraphics[width=\textwidth]{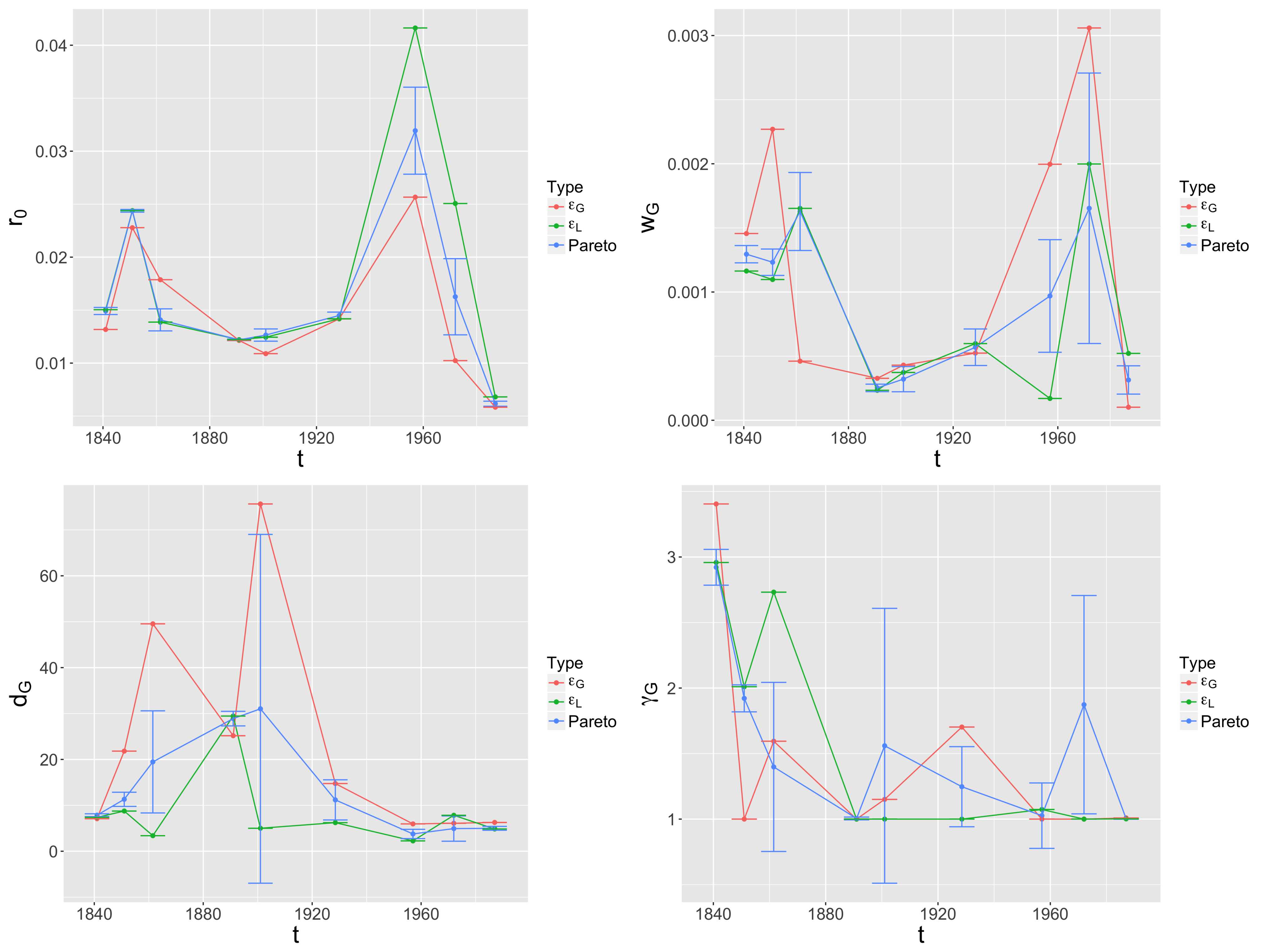}
\caption{\textbf{Calibrated parameters for gravity model only.} Each plot gives fitted values in time $t$ for each parameter: in order from left to right and top to bottom, growth rate $r_0$, gravity weight $w_G$, gravity decay $d_G$, gravity gamma $\gamma_G$. Red and Green curves correspond to best points for $\varepsilon_G$ (respectively $\varepsilon_L$), whereas the blue curves give the average value over the Pareto front with standard deviation.}
\label{fig:gravity-params}
\end{figure}

We now use the model to indirectly extract information on processes in time. Indeed, under assumption of non-stationarity, the temporal evolution of locally adjusted parameters shows the evolution of the corresponding aspect of the processes.

In a first experiment, we set $w_N=0$ and calibrate the model with four parameters on the 9 successive 20 years time windows. The optimization problem associated to model calibration does not present features allowing an easy solving (closed-form of a likelihood function, convexity or sparsity of the optimization problem), we must rely on alternative techniques to solve it. Brute force grid search is rapidly limited by the dimensionality curse. Classical methods~\citep{batty1972calibration} such as gradient descent fail because of the rather complicated shape of the optimisation landscape. Calibration using Genetic Algorithms (GA) are an efficient solution to find approximate solutions in a reasonable time. OpenMole embeds a collection of such meta-heuristics fo different purposes: \cite{schmitt2014half} demonstrate the capabilities of these methods to calibrate models of simulation. In our case, it furthermore allow to do a bi-objective calibration on $(\varepsilon_G,\varepsilon_L)$.

We use the standard steady state GA provided by OpenMole \citep{pumain2017urban}, distributed on 25 islands, with a population of 200 individuals and 100 generations. We show in Figure~\ref{fig:gravity-pareto} the calibration results on successive periods, by plotting final populations in the objective space. As expected, Pareto fronts that corresponds to compromises between the two opposite objectives are the rule. It means that the model cannot be accurate both globally and locally, and an intermediate solution has to be found. This may due to the fact that interaction range changes with city size (i.e. that terms in the potential are no longer separable), that we keep as a possible model development. The shape of the Pareto front are revealing the chaotic optimisation landscape, as for some periods such as 1921-1936 or 1962-1982 fronts are not regular and sparse. The change in shapes also translates different dynamical regimes across the periods: for 1881-1901, the quasi-vertical shape followed by an isolated front at high $\varepsilon_G$ values reveals a quasi-binary model behavior in the optimal regimes, in the sense that improving $\varepsilon_L$ under the limit is only possible through a qualitative jump at a high price for $\varepsilon_G$. The values taken by $d_G$ for periods 1892-1911 and 1921-1936 show that larger cities have longer interaction ranges, as higher values of $d_G$ give better values of $\varepsilon_G$.

We show in Figure~\ref{fig:gravity-params} the values of fitted parameters in time, averaged over the Pareto front and for best single-objective solutions. First, the two peaks patterns for $r_0$ corresponds roughly to the patterns observed in average growth rates. The evolution of $w_G$ has a similar shape but lagged by 20 years: it can be interpreted as a repercussion of endogenous growth on interaction patterns in the following years, which is consistent with an interpretation of the interaction process in terms of migrations. The values of $d_G$, with an increase until 1900 followed by a progressive decrease, is consistent with the behavior of empirical correlations commented above: the first 50 years windows in Figure~\ref{fig:ts-correlations} have a higher interaction range corresponding to flat correlation curves. Finally, the level of hierarchy $\gamma_G$ has regularly decreased, corresponding to an attenuation of the power of large cities that can be understood in terms of the progressive decentralization in France that has been fostered by the administration.

\subsection*{Unveiling Network Effects}

\begin{figure}
\centering
\includegraphics[width=\textwidth]{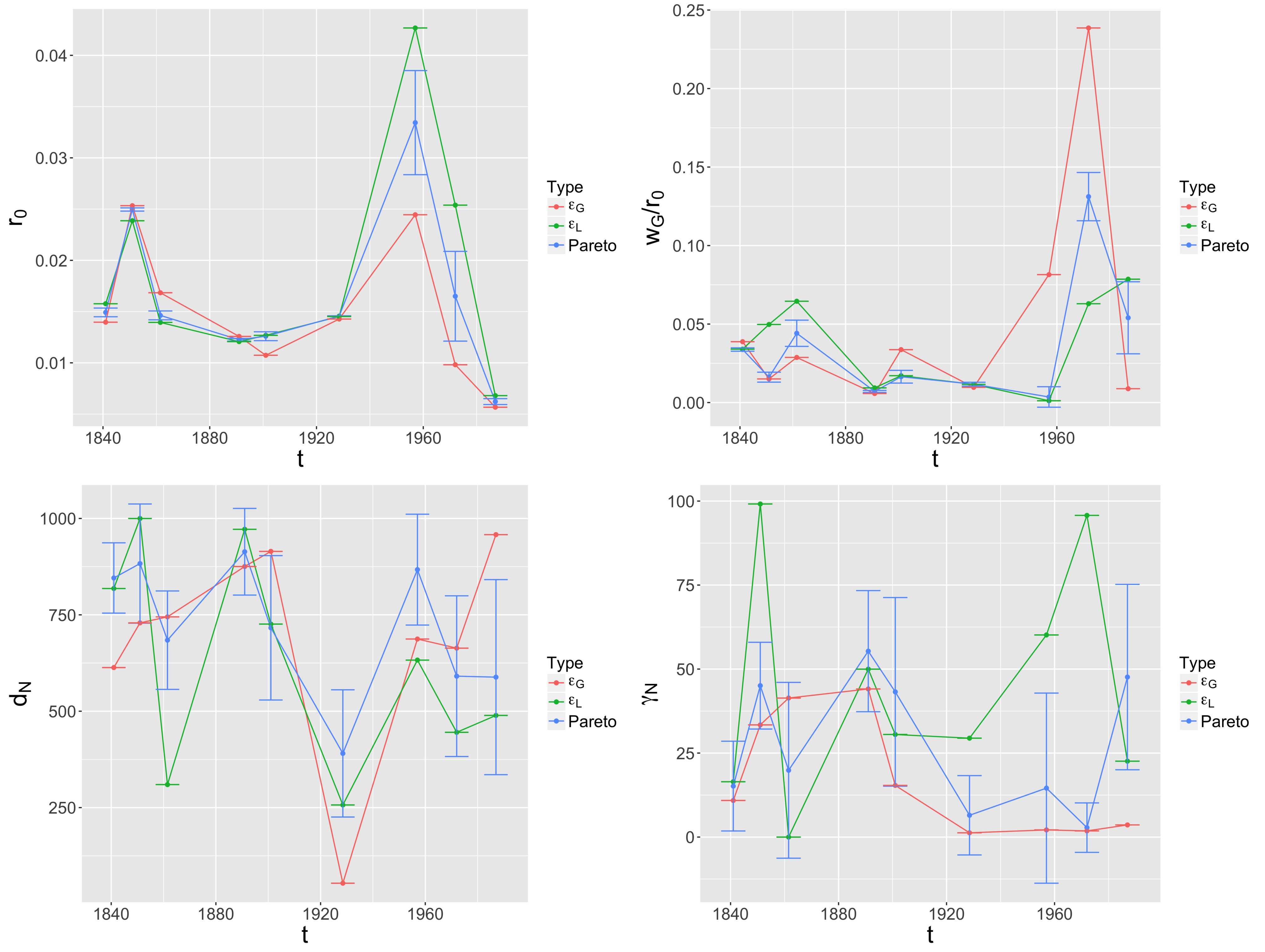}
\caption{\textbf{Calibrated parameters for the full model.} We plot values of $r_0, w_G/r_0, d_N$ and $\gamma_N$ in time $t$, for single-objective optimal points (Red and Green curves) and averaged over the Pareto front (Blue).}
\label{fig:feedback}
\end{figure}

We now turn to the calibration of the full model on successive periods, in order to interpret parameters linked to network flows and gain insight into network effects. The full calibration is done in a similar way with seven parameters being free. We plot in Figure~\ref{fig:feedback} the fitted values in time for some of these parameters.

The behavior of growth rate and of the gravity weight relative to growth rate, that is similar to the gravity model only, confirms that network effects are indeed at the second order and that endogenous growth and direct interactions are main drivers. Network effects are however not negligible, as they improve the fit as shown before in model exploration, capturing therein second order processes. The evolution of $d_N$, corresponding to the range on which network influences the territories it goes through, shows a minimum in 1921-1936 to stabilize again later, but at a value lower that past values. This could correspond to the ``tunnel effect'', when high-speed transportation does not stop in territories it goes through. Indeed, the evolution of railway has witnessed a high decrease for local lines at a date similar to the minimum, and later the emergence of specific high speed lines, explaining this lower final value. Hierarchy of flows have slightly decreased as for gravity, but are extremely high. This means that only flows between larger cities have a significant effect. This way, the model gives indirect information on the processes linked to network effects.

\subsection*{Evaluating Model Performance}

We focus in this last experiment on quantifying the ``performance'' of the model, taking into account not only its predictive capabilities, but also its structure. More precisely, we want to tackle the issue of overfitting, which has been for long recognized in the field of machine learning for example~\citep{dietterich1995overfitting}, but for which there is a lack of methods for models of simulation. We need to introduce a tool to confirm that the improvement in model fit is not only artificially due to additional parameters.

The Akaike Information Criterion (AIC) provides for statistical models for which a likelihood function is available the gain in information between two models~\citep{akaike1998information}, correcting fit improvement for number of parameters. Similar methods include the Bayesian Information Criterion (BIC), which relies on slightly different assumptions. \cite{biernacki2000assessing} propose an integrated likelihood as a generalization of these criteria for unsupervised classification. \cite{2017arXiv170108673P} show that in the case of selecting the number of states in Hidden Markov Models, real cases induces too much pitfalls for standard methods to work robustly, and suggest pragmatic selection based on their results and expert judgement. In our case, the problem is that it is not even possible to define the criteria.

The method we propose is based on the intuitive idea of approaching models of simulation by statistical models and using the corresponding AIC under certain validity conditions. \cite{2017arXiv170609773B} use a similar trick of considering the models as black boxes and approaching them to gain insights, in their case to extract interpretable structure as decision trees.

Let $(X,Y)$ be the data and observations. We consider computational models as functions $(X,\alpha_k) \mapsto M_{\alpha_k}^{(k)}(X)$ mapping data values to a random variable. What is seen as data and parameters is somehow arbitrary but is separated in the formulation as corresponding dimensions will have different roles. We assume that the models have been fitted to data in the sense that an heuristic has been used to find an approximate optimal solution $\alpha^{\ast}_k = \textrm{argmin}_{\alpha_k}\norm{M_{\alpha_k}^{(k)}(X) - Y}$, and we write $\varepsilon_k = \norm{M_{\alpha_k}^{(k)}(X) - Y}^2$ the corresponding mean-square error. For each optimized computational model, a statistical model $S^{(k)}$ with the same degree of freedom can be fitted on a set of realizations: $M^{(k)}_{\alpha^{\ast}_k}(X) = S^{(k)} (X)$, with an error $s_k = \norm{M_{\alpha^{\ast}_k}^{(k)}(X) - S^{(k)}(X)}^2$.

We follow the intuition that if statistical models are good approximations of the simulation models compared to models discrepancy to reality, namely $s_k \ll \varepsilon_k$, then the gain of information between the two should mostly capture the gain of information between simulation models\footnote{There is to the best of our knowledge no way to establish this relation formally. According to \cite{navakatikyan2007model}, we have $AIC^{(k)} = n \log{\left(s_k/n\right)}+2K_k$ if $n$ is the number of data points and $K_k$ the number of parameters for model $k$. It follows that $I = n \Delta \log{\left(s_k/\varepsilon_k\right)} + n \Delta \log{\left(\varepsilon_k/n\right)} + 2 \Delta K_k$. The two last terms can however not be related to the gain of information in simulation models as these have no likelihood function.}. We define therefore an \emph{Empirical AIC} measure between two simulation models by

\begin{equation}
I\left( M^{(1)}, M^{(2)}\right) = \Delta AIC \left[S^{(1)},S^{(2)}\right]
\end{equation}

In practice we calibrate the gravity only model and the full model on the full time span, and choose two intermediate solutions giving $M^{(1)}$ at $r_0=0.0133, d_G = 4.02e12, w_G = 1.28e-4, \gamma_G = 3.82$ with $\varepsilon_G=31.2375,\varepsilon_L=302.89$ and the full model $M^{(2)}$ at $r_0=0.0128, d_G = 8.43e14, w_G = 1.230e-4, \gamma_G = 3.81, w_N=0.60, d_N=7.47e14, \gamma_N = 1.15$ with $\varepsilon_G=31.2366,\varepsilon_L=302.93$. It is not clear how the empirical method is sensitive to the type of statistical model used. We use therefore several models for robustness, each time with the corresponding number of parameters (4 for the first and 7 for the second model): a polynomial model of the form $a_0 + \sum_{i>0} a_i X^i$, a mixture of logarithm and polynomial as $a_0 + a_1 \ln X + \sum_{i>1} a_i X^i$ and a generalized polynomial with real power coefficients that are optimized for model fit using a genetic algorithm $a_0 + \sum_{i>0} a_i X^{\alpha_i}$. We fit the statistical models using successive years as different realizations.

Results  are shown in Table~\ref{tab:empiricalaic}. We give the value of $s_k / \varepsilon_k$ and the $\Delta AIC$. We also provide the $\Delta BIC$, that correspond to the bayesian criterion differences on statistical models, to check the robustness regarding the information criterion used. We find a positive value for 5 criteria out of 6, confirming that information gain is indeed positive. The gain decreases when statistical model fit improves, and only the BIC for the optimized model fails to show an improvement. The assumption of negligible errors is always verified as the rate stays around 1\%. This approach is of course preliminary and further work should be done for a more systematic testing and a more robust justification of the method. It suggests however that fit improvement in the model of simulation are effective, and that the model reveals therefore network effects.

\begin{table}[ht]
\small\sf\centering
\caption{\textbf{Empirical AIC results.}}\label{tab:empiricalaic}
\begin{tabular}{|l|l|l|l|l|l}
\hline
Statistical Model & $M^{(1)}$ Relative fit & $M^{(2)}$ Relative fit & $\Delta AIC$ & $\Delta BIC$\\
\hline
Polynomial & 0.01438 & 0.01415 & 19.59 & 3.65\\
Log-polynomial & 0.01565  & 0.01435 & 125.37 & 109.43\\
Generalized Polynomial & 0.01415  & 0.01399 & 11.70 & -4.23\\
\hline
\end{tabular}
\end{table}

\section*{Discussion}

\subsection*{Multi-modeling}

We obtain for most of periods and calibration points large values of $\gamma_N$ in the calibration of the full model, what suggests that a model with only major flows could be a possible alternative. An additional parameter determining the cutoff value for flows would have to be introduced, and to what extent this variation improves model performance should be tested in future developments. More generally, this raises the issue of more systematic multi-modeling approaches and of benchmarking models. \cite{cottineau2015modular} introduce a modular framework to compare concurrent hypothesis to explain urban growth. In the same spirit, an immediate extension of our work would be to test different functional specifications and benchmark more diverse network processes, such as the diffusion of innovation as done by \cite{favaro2011gibrat} or the creation of social ties \citep{frasco2014spatially}.

\subsection*{Theoretical implications}

Our results support the hypothesis that physical transportation networks are necessary to explain the morphogenesis of territorial systems, in the sense that some aspects are fully contained within networks and cannot be approximated by abstract proxies. We showed indeed on a relatively simple case that the integration of physical networks into some models effectively increase their explanative power even when controlling for overfitting. This can be understood as a direction to expand Pumain's Evolutive Urban Theory~\citep{pumain1997pour}. This theory considers networks as carriers of interactions in systems of cities but does not put a particular emphasis on their physical aspect and the possible spatial patterns resulting from it such as bifurcations or network induced differentiations. The development of a sub-theory focusing on these aspects is an interesting direction suggested by our empirical and modeling results.

\subsection*{Urban System Specificity}

The model has not yet been tested on other urban systems and other temporalities, and further work should investigate which conclusions we obtained here are specific to the French urban system on this period, and which are more general to any system of cities. Applying the model to other system of cities also recalls the difficulty of defining urban systems. In our case, a strong bias should arise by considering France only, as Lille must be highly influenced by Brussels for example. The extent and scale of such models is always a delicate question. We rely here on the administrative coherence and the consistence of the database, but sensitivity to system definition and extent should also be further tested.

\subsection*{Towards co-evolutive models}

Our focus on network effects remains quite limited since (i) we do not consider an effective infrastructure but abstract flows only, and (ii) we do not take into account the possible network evolution, due to technical progresses~\citep{bretagnolle2000long} and infrastructure growth in time. An interesting development would be first the application of our model with real network data, using effective distance matrices in time, computed e.g. with the train network used by~\cite{thevenin2013mapping}.

Then, allowing the network to dynamically evolve in time, as a function of flows, would yield a model of co-evolution between cities and transportation networks for a system of cities. This co-evolution has been highlighted empirically by~\cite{bretagnolle:tel-00459720}. This kind of model is not frequent, and \cite{schmitt2014modelisation} provides with the SimpopNet model one of the few examples. It is shown by \cite{raimbault2016models} that disciplinary compartmentalization may be at the origin of the relative absence of such type of models in the literature. Indeed, it would imply to include heterogenous processes such as economic rules to drive network growth, that are quite far from the approach taken. It would however allow to investigate to what extent the refinement of network spatial structure and network dynamics can improve the explanation of urban system dynamics.

\section*{Conclusion}

We have introduced a spatial model of growth for a system of cities at the macroscopic scale, including second order network effects among endogenous growth and direct interaction growth drivers. The model is parametrized on real data for the French city system between 1831 and 1999. The calibration of the model in time provides interpretations for the evolution of processes of interaction within the system of cities. We furthermore show that the model effectively unveils network effects by controlling for overfitting. This work paves the way for more complicated models with dynamical networks, that would capture the co-evolution between transportation network and territories.

\section*{Acknowledgments}
 Results obtained in this paper were computed on the vo.complex-system.eu virtual organization of the European Grid Infrastructure ( http://www.egi.eu ). We thank the European Grid Infrastructure and its supporting National Grid Initiatives (France-Grilles in particular) for providing the technical support and infrastructure.


\begin{thebibliography}{51}
\providecommand{\natexlab}[1]{#1}
\providecommand{\url}[1]{\texttt{#1}}
\providecommand{\urlprefix}{URL }
\expandafter\ifx\csname urlstyle\endcsname\relax
  \providecommand{\doi}[1]{DOI:\discretionary{}{}{}#1}\else
  \providecommand{\doi}{DOI:\discretionary{}{}{}\begingroup
  \urlstyle{rm}\Url}\fi

\bibitem[{Akaike(1998)}]{akaike1998information}
Akaike H (1998) Information theory and an extension of the maximum likelihood
  principle.
\newblock In: \emph{Selected Papers of Hirotugu Akaike}. Springer, pp.
  199--213.

\bibitem[{Andersson et~al.(2006)Andersson, Frenken and
  Hellervik}]{andersson2006complex}
Andersson C, Frenken K and Hellervik A (2006) A complex network approach to
  urban growth.
\newblock \emph{Environment and Planning A} 38(10): 1941.

\bibitem[{Baptiste(1999)}]{baptiste1999interactions}
Baptiste H (1999) \emph{Interactions entre le syst{\`e}me de transport et les
  syst{\`e}mes de villes: perspective historique pour une mod{\'e}lisation
  dynamique spatialis{\'e}e}.
\newblock PhD Thesis, Centre d'{\'e}tudes sup{\'e}rieures de l'am{\'e}nagement
  (Tours).

\bibitem[{{Bastani} et~al.(2017){Bastani}, {Kim} and
  {Bastani}}]{2017arXiv170609773B}
{Bastani} O, {Kim} C and {Bastani} H (2017) {Interpretability via Model
  Extraction}.
\newblock \emph{arXiv preprint arXiv:1706.09773} .

\bibitem[{Batty and Mackie(1972)}]{batty1972calibration}
Batty M and Mackie S (1972) The calibration of gravity, entropy, and related
  models of spatial interaction.
\newblock \emph{Environment and Planning A} 4(2): 205--233.

\bibitem[{Bedau(2002)}]{bedau2002downward}
Bedau M (2002) Downward causation and the autonomy of weak emergence.
\newblock \emph{Principia: an international journal of epistemology} 6(1):
  5--50.

\bibitem[{Bettencourt et~al.(2008)Bettencourt, Lobo and
  West}]{bettencourt2008large}
Bettencourt LM, Lobo J and West GB (2008) Why are large cities faster?
  universal scaling and self-similarity in urban organization and dynamics.
\newblock \emph{The European Physical Journal B-Condensed Matter and Complex
  Systems} 63(3): 285--293.

\bibitem[{Biernacki et~al.(2000)Biernacki, Celeux and
  Govaert}]{biernacki2000assessing}
Biernacki C, Celeux G and Govaert G (2000) Assessing a mixture model for
  clustering with the integrated completed likelihood.
\newblock \emph{IEEE transactions on pattern analysis and machine intelligence}
  22(7): 719--725.

\bibitem[{Bigotte et~al.(2010)Bigotte, Krass, Antunes and
  Berman}]{bigotte2010integrated}
Bigotte JF, Krass D, Antunes AP and Berman O (2010) Integrated modeling of
  urban hierarchy and transportation network planning.
\newblock \emph{Transportation Research Part A: Policy and Practice} 44(7):
  506--522.

\bibitem[{Bretagnolle(2009)}]{bretagnolle:tel-00459720}
Bretagnolle A (2009) \emph{{Villes et r{\'e}seaux de transport : des
  interactions dans la longue dur{\'e}e, France, Europe, {\'E}tats-Unis}}.
\newblock Hdr, Universit{\'e} Panth{\'e}on-Sorbonne - Paris I.

\bibitem[{Bretagnolle et~al.(2000)Bretagnolle, Mathian, Pumain and
  Rozenblat}]{bretagnolle2000long}
Bretagnolle A, Mathian H, Pumain D and Rozenblat C (2000) Long-term dynamics of
  european towns and cities: towards a spatial model of urban growth.
\newblock \emph{Cybergeo: European Journal of Geography} .

\bibitem[{Bretagnolle and Pumain(2010)}]{bretagnolle2010comparer}
Bretagnolle A and Pumain D (2010) Comparer deux types de syst{\`e}mes de villes
  par la mod{\'e}lisation multi-agents.
\newblock \emph{Qu'appelle t-on aujourd'hui les sciences de la complexit{\'e}?
  Langages, r{\'e}seaux, march{\'e}s, territoires} : 271--299.

\bibitem[{Chang(2006)}]{chang2006models}
Chang JS (2006) Models of the relationship between transport and land-use: A
  review.
\newblock \emph{Transport Reviews} 26(3): 325--350.

\bibitem[{Collischonn and Pilar(2000)}]{collischonn2000direction}
Collischonn W and Pilar JV (2000) A direction dependent least-cost-path
  algorithm for roads and canals.
\newblock \emph{International Journal of Geographical Information Science}
  14(4): 397--406.

\bibitem[{Cottineau(2014)}]{cottineau2014evolution}
Cottineau C (2014) \emph{L'{\'e}volution des villes dans l'espace
  post-sovi{\'e}tique. Observation et mod{\'e}lisations.}
\newblock PhD Thesis, Universit{\'e} Paris 1 Panth{\'e}on-Sorbonne.

\bibitem[{Cottineau et~al.(2015)Cottineau, Reuillon, Chapron, Rey-Coyrehourcq
  and Pumain}]{cottineau2015modular}
Cottineau C, Reuillon R, Chapron P, Rey-Coyrehourcq S and Pumain D (2015) A
  modular modelling framework for hypotheses testing in the simulation of
  urbanisation.
\newblock \emph{Systems} 3(4): 348--377.

\bibitem[{Dietterich(1995)}]{dietterich1995overfitting}
Dietterich T (1995) Overfitting and undercomputing in machine learning.
\newblock \emph{ACM computing surveys (CSUR)} 27(3): 326--327.

\bibitem[{Favaro and Pumain(2011)}]{favaro2011gibrat}
Favaro JM and Pumain D (2011) Gibrat revisited: An urban growth model
  incorporating spatial interaction and innovation cycles.
\newblock \emph{Geographical Analysis} 43(3): 261--286.

\bibitem[{Frasco et~al.(2014)Frasco, Sun, Rozenfeld and
  Ben-Avraham}]{frasco2014spatially}
Frasco GF, Sun J, Rozenfeld HD and Ben-Avraham D (2014) Spatially distributed
  social complex networks.
\newblock \emph{Physical Review X} 4(1): 011008.

\bibitem[{Gabaix(1999)}]{gabaix1999zipf}
Gabaix X (1999) Zipf's law for cities: an explanation.
\newblock \emph{Quarterly journal of Economics} : 739--767.

\bibitem[{Glaeser(2011)}]{glaeser2011triumph}
Glaeser E (2011) \emph{Triumph of the city: How our greatest invention makes us
  richer, smarter, greener, healthier, and happier}.
\newblock Penguin.

\bibitem[{Gu{\'e}rin-Pace and Pumain(1990)}]{guerin1990150}
Gu{\'e}rin-Pace F and Pumain D (1990) 150 ans de croissance urbaine.
\newblock \emph{Economie et statistique} 230(1): 5--16.

\bibitem[{Krugman(1998)}]{krugman1998space}
Krugman P (1998) Space: the final frontier.
\newblock \emph{The Journal of Economic Perspectives} 12(2): 161--174.

\bibitem[{Mantegna and Stanley(1999)}]{mantegna1999introduction}
Mantegna RN and Stanley HE (1999) \emph{Introduction to econophysics:
  correlations and complexity in finance}.
\newblock Cambridge university press.

\bibitem[{Marchionni(2004)}]{marchionni2004geographical}
Marchionni C (2004) Geographical economics versus economic geography: towards a
  clarification of the dispute.
\newblock \emph{Environment and Planning A} 36(10): 1737--1753.

\bibitem[{Masucci et~al.(2013)Masucci, Serras, Johansson and
  Batty}]{masucci2013gravity}
Masucci AP, Serras J, Johansson A and Batty M (2013) Gravity versus radiation
  models: On the importance of scale and heterogeneity in commuting flows.
\newblock \emph{Physical Review E} 88(2): 022812.

\bibitem[{Navakatikyan(2007)}]{navakatikyan2007model}
Navakatikyan MA (2007) A model for residence time in concurrent variable
  interval performance.
\newblock \emph{Journal of the experimental analysis of behavior} 87(1):
  121--141.

\bibitem[{Nitsch(2005)}]{nitsch2005zipf}
Nitsch V (2005) Zipf zipped.
\newblock \emph{Journal of Urban Economics} 57(1): 86--100.

\bibitem[{{Pohle} et~al.(2017){Pohle}, {Langrock}, {van Beest} and
  {Schmidt}}]{2017arXiv170108673P}
{Pohle} J, {Langrock} R, {van Beest} F and {Schmidt} NM (2017) {Selecting the
  Number of States in Hidden Markov Models - Pitfalls, Practical Challenges and
  Pragmatic Solutions}.
\newblock \emph{arXiv preprint arXiv:1701.08673} .

\bibitem[{Pumain(1997)}]{pumain1997pour}
Pumain D (1997) Pour une th{\'e}orie {\'e}volutive des villes.
\newblock \emph{Espace g{\'e}ographique} 26(2): 119--134.

\bibitem[{Pumain(2012{\natexlab{a}})}]{pumain2012multi}
Pumain D (2012{\natexlab{a}}) Multi-agent system modelling for urban systems:
  The series of simpop models.
\newblock In: \emph{Agent-based models of geographical systems}. Springer, pp.
  721--738.

\bibitem[{Pumain(2012{\natexlab{b}})}]{pumain2012urban}
Pumain D (2012{\natexlab{b}}) Urban systems dynamics, urban growth and scaling
  laws: The question of ergodicity.
\newblock In: \emph{Complexity Theories of Cities Have Come of Age}. Springer,
  pp. 91--103.

\bibitem[{Pumain et~al.(2009)Pumain, Paulus and
  Vacchiani-Marcuzzo}]{pumain2009innovation}
Pumain D, Paulus F and Vacchiani-Marcuzzo C (2009) Innovation cycles and urban
  dynamics.
\newblock \emph{Complexity perspectives in innovation and social change} :
  237--260.

\bibitem[{Pumain et~al.(2006)Pumain, Paulus, Vacchiani-Marcuzzo and
  Lobo}]{pumain2006evolutionary}
Pumain D, Paulus F, Vacchiani-Marcuzzo C and Lobo J (2006) An evolutionary
  theory for interpreting urban scaling laws.
\newblock \emph{Cybergeo: European Journal of Geography} .

\bibitem[{Pumain and Reuillon(2017{\natexlab{a}})}]{pumain2017simpoplocal}
Pumain D and Reuillon R (2017{\natexlab{a}}) The simpoplocal model.
\newblock In: \emph{Urban Dynamics and Simulation Models}. Springer, pp.
  21--35.

\bibitem[{Pumain and Reuillon(2017{\natexlab{b}})}]{pumain2017urban}
Pumain D and Reuillon R (2017{\natexlab{b}}) \emph{Urban Dynamics and
  Simulation Models}.
\newblock Springer International.

\bibitem[{Pumain and Riandey(1986)}]{pumain1986fichier}
Pumain D and Riandey B (1986) Le fichier de l'ined.
\newblock \emph{Espace, populations, soci{\'e}t{\'e}s} 4(2): 269--277.

\bibitem[{Pumain and Sanders(2013)}]{pumain2013theoretical}
Pumain D and Sanders L (2013) Theoretical principles in interurban simulation
  models: a comparison.
\newblock \emph{Environment and Planning A} 45(9): 2243--2260.

\bibitem[{Raimbault(2017)}]{raimbault2016models}
Raimbault J (2017) Models coupling urban growth and transportation network
  growth: An algorithmic systematic review approach.
\newblock \emph{Plurimondi} (17).

\bibitem[{Reuillon et~al.(2013)Reuillon, Leclaire and
  Rey-Coyrehourcq}]{reuillon2013openmole}
Reuillon R, Leclaire M and Rey-Coyrehourcq S (2013) Openmole, a workflow engine
  specifically tailored for the distributed exploration of simulation models.
\newblock \emph{Future Generation Computer Systems} 29(8): 1981--1990.

\bibitem[{Rozenfeld et~al.(2008)Rozenfeld, Rybski, Andrade, Batty, Stanley and
  Makse}]{rozenfeld2008laws}
Rozenfeld HD, Rybski D, Andrade JS, Batty M, Stanley HE and Makse HA (2008)
  Laws of population growth.
\newblock \emph{Proceedings of the National Academy of Sciences} 105(48):
  18702--18707.

\bibitem[{Rybski et~al.(2013)Rybski, Ros and Kropp}]{rybski2013distance}
Rybski D, Ros AGC and Kropp JP (2013) Distance-weighted city growth.
\newblock \emph{Physical Review E} 87(4): 042114.

\bibitem[{Sanders(1992)}]{sanders1992systeme}
Sanders L (1992) \emph{Syst{\`e}me de villes et synerg{\'e}tique}.
\newblock Economica.

\bibitem[{Sanders et~al.(1997)Sanders, Pumain, Mathian, Gu{\'e}rin-Pace and
  Bura}]{sanders1997simpop}
Sanders L, Pumain D, Mathian H, Gu{\'e}rin-Pace F and Bura S (1997) Simpop: a
  multiagent system for the study of urbanism.
\newblock \emph{Environment and Planning B} 24: 287--306.

\bibitem[{Schmitt(2014)}]{schmitt2014modelisation}
Schmitt C (2014) \emph{Mod{\'e}lisation de la dynamique des syst{\`e}mes de
  peuplement: de SimpopLocal {\`a} SimpopNet.}
\newblock PhD Thesis, Paris 1.

\bibitem[{Schmitt et~al.(2015)Schmitt, Rey-Coyrehourcq, Reuillon and
  Pumain}]{schmitt2014half}
Schmitt C, Rey-Coyrehourcq S, Reuillon R and Pumain D (2015) Half a billion
  simulations: Evolutionary algorithms and distributed computing for
  calibrating the simpoplocal geographical model.
\newblock \emph{Environment and Planning B: Planning and Design} 42(2):
  300--315.

\bibitem[{Storper and Scott(2009)}]{storper2009rethinking}
Storper M and Scott AJ (2009) Rethinking human capital, creativity and urban
  growth.
\newblock \emph{Journal of economic geography} 9(2): 147--167.

\bibitem[{Taylor(2016)}]{taylor2016polymath}
Taylor PJ (2016) A polymath in city studies.
\newblock In: \emph{Sir Peter Hall: Pioneer in Regional Planning, Transport and
  Urban Geography}. Springer, pp. 11--20.

\bibitem[{Th{\'e}venin et~al.(2013)Th{\'e}venin, Schwartz and
  Sapet}]{thevenin2013mapping}
Th{\'e}venin T, Schwartz R and Sapet L (2013) Mapping the distortions in time
  and space: The french railway network 1830--1930.
\newblock \emph{Historical Methods: A Journal of Quantitative and
  Interdisciplinary History} 46(3): 134--143.

\bibitem[{West(2017)}]{west2017scale}
West G (2017) \emph{Scale: The Universal Laws of Growth, Innovation,
  Sustainability, and the Pace of Lifein Organisms, Cities, Economies, and
  Companies}.
\newblock Penguin.

\bibitem[{Xie and Levinson(2009)}]{xie2009modeling}
Xie F and Levinson D (2009) Modeling the growth of transportation networks: A
  comprehensive review.
\newblock \emph{Networks and Spatial Economics} 9(3): 291--307.

\end{thebibliography}
\end{document}